\documentstyle[11pt]{article}
\begin{document}
\topmargin -15mm
\textheight 225mm
\baselineskip14pt

%%%%%%%If you do not have the msbm fonts, delete the following 4 lines
%\font\mybb=msbm10 at 12pt
%\def\bb#1{\hbox{\mybb#1}}
%\def\Z {\bb{Z}}
%\def\R {\bb{R}}
%%%%%%%%%%%%
%%%and replace with the following 2 lines (without %)
\def\Z {Z}
\def\R {R}
%%%%%%%%%%
 \def\unit{\hbox to 3.3pt{\hskip1.3pt \vrule height 7pt width .4pt \hskip.7pt
\vrule height 7.85pt width .4pt \kern-2.4pt
\hrulefill \kern-3pt
\raise 4pt\hbox{\char'40}}}
\def\II{{\unit}}
\def\cM {{\cal{M}}}
\def\half{{\textstyle {1 \over 2}}}

\def\threeint{3$\perp$3$\perp$3$\perp$3}
\def\fourint{4$\perp$4$\perp$4$\perp$4}
\def\fiveint{5$\perp$5$\perp$5$\perp$5}
\def\foursixint{4$\perp$4$\perp$6$\perp$6}
\def\foureightint{4$\perp$4$\perp$4$\perp$8}
\def\twosixint{2$\perp$6$\perp$6$\perp$6}
\def\threefivesevenint{3$\perp$5$\perp$5$\perp$7}

\pagestyle{empty}
\rightline{UG-7/96}
\rightline{MRI-PHY/96-27}
\rightline{September 1996}
\rightline{hep-th/9609056}
\vspace{1.8truecm}
\centerline{\bf Four-dimensional High-Branes as Intersecting $D$-Branes}
\vspace{1.8truecm}
\centerline{E. BERGSHOEFF, M. DE ROO}
\vspace{.2truecm}
\centerline{\it Institute for Theoretical Physics}
\centerline{\it University of Groningen, Nijenborgh 4}
\centerline{\it 9747 AG Groningen}
\centerline{\it The Netherlands}
\vspace{0.5truecm}
\centerline{and}
\vspace{0.5truecm}
\centerline{S. PANDA}
\vspace{.2truecm}
\centerline{\it Mehta Research Institute of Mathematics \& Mathematical 
   Physics}
\centerline{\it 10 Kasturba Gandhi Marg}
\centerline{\it Allahabad 211002}
\centerline{\it India}
\vspace{1.8truecm}
\centerline{ABSTRACT}
\vspace{.5truecm}
We show that a class of extremal 
four-dimensional supersymmetric \hfill\break
``high\-branes'', i.e.~string and
domain wall solutions, can be interpreted as intersections of four
ten-dimensional Dirichlet branes. These $d=4$ solutions are
related, via $T$-duality in ten dimensions, to the four-dimensional extremal 
Maxwell/scalar black holes that are characterized by a scalar coupling 
parameter $a$ with $a=0,1/\sqrt 3, 1, \sqrt 3$.

\vfill\eject
\pagestyle{plain}
 
\noindent{\bf 1. Introduction}
\vspace{.5cm}

The construction of solutions to the Einstein equations, including
 matter in the form of scalar fields and gauge fields, has advanced
 rapidly in recent years due to the developments in string theory.
 Solution generating transformations, which follow from the
 $T$ and $S$ dualities of string theory, produce quite intricate 
 $p$-brane solutions 
 from relatively simple ones\footnote{For a review of fundamental
 and solitonic solutions, see, e.g.,
 \cite{du/kh/lu}.}. 
 
Some extremal solutions can be understood in terms of 
bound states \cite{ra,cv/ts,kh/lu/po,du/ra} or as intersections
 of $D$-branes ($M$-branes) in ten (eleven) dimensions
 \cite{pa/to,ts1,be/be/ja,ga/ka/tr,ba/la,be/be}. 
 The construction of lower dimensional solutions in terms of
 ten-dimensional ones suggests a possible $p$-brane classification 
 scheme.  In this letter we provide further support for this
 point of view by showing that not only black holes ($p=0$) but also
 a class of extremal string ($p=1$) and domain wall ($p=2$) solutions 
 in four dimensions can be understood as intersections of ten-dimensional
$D$-branes. Furthermore, we show that all $p=0,1,2$ solutions
are related to each other via $T$-duality in ten dimensions, thereby
providing a unifying picture for all these $d=4$ solutions.

The black hole solutions to string effective actions in diverse
 dimensions have drawn the most attention in this context\footnote{For a 
 recent overview, see \cite{ts2}.}.
 The four
 supersymmetric, single charge, extreme dilaton black hole solutions in 
 four dimensions can
 all be understood in terms of the intersection of Dirichlet 3-branes
 in ten dimensions. 
 This intersection is described by a ten-dimensional
 metric containing four independent harmonic functions \cite{ts1,ba/la}.
 The single charge black hole solutions that are characterized by a 
scalar coupling 
parameter $a$ with $a=0,1/\sqrt 3, 1, \sqrt 3$ can be obtained by the
 identification or absence of some of these harmonic functions.
 Black hole solutions in terms of four harmonic functions were obtained
 earlier in \cite{ra,cv/ts}, and generalized to $p>0$-branes in 
 \cite{kh/lu/po,du/ra}. These solutions generically contain several scalar
 fields, and are generalizations of a class of solutions
discussed in \cite{lu/po}.
 
General $p$-branes in $d$ dimensions can be divided in two
categories: (i) ``low-branes'' ($p=0,\ldots, d-4$), which couple
 to the fundamental (or dual) gauge fields of the underlying 
supergravity theory and (ii) ``high-branes''
($p=d-3,d-2$) which, in contrast to the low-branes,
are not asymptotically flat. The $p$-branes whose charge is carried by a
 NS-NS gauge field have been understood for some time. The $p$-branes
 coupling to RR fields or their duals are represented by the so-called 
Dirichlet
 ($D$)-branes \cite{po,po/ch/jo}. In ten dimensions $D$-branes
 exist for all $p=0,\ldots,9$. They correspond to solutions of the
 IIA (IIB) supergravity theories for $p$ even (odd), with the following 
 string frame metric and dilaton:
\begin{eqnarray}
  ds^2_{S,10} &=& H^{-1/2}dx^2_{(p+1)} - H^{1/2}dx^2_{(9-p)}\,, \nonumber\\
\label{101}
  e^{2\phi}   &=& H^{(3-p)/2}\,,
\end{eqnarray}
where $H$ is a harmonic function depending on the $9-p$ transverse
 coordinates. These $D$-branes are related by $T$-duality, which 
 turns a $p$-brane into a $p+1$-brane solution, by \cite{be/hu/or,ber/roo}:
\begin{equation}
\label{102}
 g'_{\mu\nu} = g_{\mu\nu}\,,\quad
 g'_{xx} = 1/ g_{xx}\,,\quad  e^{2\phi'} = e^{2\phi}|g_{xx}|^{-1}\, ,\nonumber
\end{equation}
where $x$ is one of the transverse coordinates, and it is understood that
 $H$ is independent of $x$. This duality transformation acts between
 the IIA and IIB theories.

The four-dimensional black holes can be understood as an intersection of four
 D3-branes: \threeint\ \cite{ts1,ba/la}. We will
 construct the four-dimensional high-branes (strings and domain walls)
 in terms of similar intersections of $d=10$ D-branes. That will be done
 in the next section. In section 3 we will discuss our results.

\vspace{.5cm}
\noindent{\bf 2. Strings and domain walls as $D$-brane intersections}
\vspace{.5truecm}

The metric and the dilaton of intersecting $D$-branes in ten dimensions 
 have a structure involving products of the individual harmonic
 functions. The possible intersections of two $D$-branes were 
 investigated in \cite{be/be/ja,ga/ka/tr}. If a $p+r$ and a $p+s$ brane
 intersect over a $p$-brane, the metric has an overall 
 world volume part ($p+1$ coordinates), and at most $9-p-r-s$
 overall transverse coordinates. Generically, an
 intersection of $N$ $D$-branes leads to a configuration with
 $32/2^N$ unbroken supersymmetry generators\footnote{Note that 
 $p,r,s$ have to satisfy certain consistency conditions, and that
 arbitrary $p,r,s$ do not always lead to solutions of the equations 
 of motion. Also, some configurations may have no unbroken supersymmetry
 generators \cite{po/ch/jo}.}.

The \threeint\ intersection has the property that each pair of 
 3-branes intersect over a 1-brane, and that these 1-branes intersect
 over a 0-brane. There are then one overall world-volume  (time--)
coordinate
 with metric component $(H_1H_2H_3H_4)^{-1/2}$
 and three overall transverse coordinates
 with metric components $(H_1H_2H_3H_4)^{1/2}$. The harmonic functions
 depend on the overall transverse coordinates only.
 Reduction over the 
 remaining six dimensions results in a four dimensional solution 
 involving four intersecting 0-branes, and produces the solution of
 \cite{ra,cv/ts}, involving three scalar fields\footnote{The ten-dimensional
 dilaton vanishes for a 3-brane solution, which implies that the
 \threeint\ intersection has one scalar less than the cases discussed 
 below.}.

Using $T$-duality over the overall transverse directions, we can
 turn the \threeint\ intersection into a \fourint\ and \fiveint\ 
 intersection which intersect over a 1-brane and 2-brane, respectively.
The reduction of these $T$-dual solutions to four dimensions naturally
results into string and
 domain wall solutions involving four independent harmonic functions.
 Given the fact that \threeint\ is a solution in $d=10$ (which indirectly
 follows from \cite{ra,cv/ts}), and that $T$-duality and reduction 
 does not change this 
 property, the four-dimensional strings and domain walls, 
 involving four scalars and
 four vector fields, also satisfy the equations of motion.

The ten-dimensional solution with four intersecting $(p+3)$-branes
$(p=0,1,2$) has the following  string frame metric and dilaton:
\begin{eqnarray}
   ds^2_{S,10} &=& {1\over\sqrt{H_1H_2H_3H_4}}dx^2_{(p+1)}
                  -\sqrt{H_1H_2H_3H_4}dx^2_{(3-p)} \nonumber\\
  &&
      -\sqrt{{H_3H_4\over H_1H_2}}dx^2_4 -\sqrt{{H_2H_4\over H_1H_3}}d^2x_5
      -\sqrt{{H_2H_3\over H_1H_4}}dx^2_6
     \nonumber\\
\label{203}
  &&  -\sqrt{{H_1H_2\over H_3H_4}}dx^2_7
     -\sqrt{{H_1H_3\over H_2H_4}}dx^2_8 -\sqrt{{H_1H_4\over H_2H_3}}dx^2_9
     \,,\\
\label{204}
  e^{2\phi} &=& (H_1H_2H_3H_4)^{-p/2}\,.
\end{eqnarray}
The solutions for different
 $p$ are related by $T$-duality in one of the $(3-p)$ overall
transverse directions. Besides the metric and dilaton, there is also
a nontrivial $(p+5)$-form field-strength whose components are
\begin{eqnarray}
\label{new}
 F_{(p+5)} &=& dx_0\wedge \cdots dx_p \wedge 
  \biggl ( dx_4\wedge dx_5\wedge dx_6\wedge dH_1^{-1} +
   \nonumber\\
  && +\ dx_4\wedge dx_8\wedge dx_9 \wedge dH_2^{-1}
  +\ dx_5\wedge dx_7\wedge dx_9\wedge dH_3^{-1} +\\
  &&\quad
  +\  dx_6\wedge dx_7\wedge dx_8\wedge dH_4^{-1}\biggr )\, .\nonumber
\end{eqnarray}

We can go to the Einstein frame by a rescaling of the metric:
\begin{equation}
\label{205}
    g_{E,10} = e^{-\phi/2} g_{S,10}\,.
\end{equation}
The action in $d=10$ then takes on the form\footnote{For even $p$ 
 this is  IIB supergravity, for which there is no action. Here
 we employ the pseudo-action defined in \cite{be/bo/or}, which
 one can freely use in dimensional reduction.}:
\begin{equation}
\label{206}
   {\cal L}_{E,10} = \sqrt{|g|}\left[ R + {1\over 2}(\partial\phi)^2
         + {(-1)^p\over 2(p+5)!}
      e^{-p\phi/2}F^2_{(p+5)} \right]\,.
\end{equation}
In the reduction to four dimensions we parametrize the metric as 
follows: 
\begin{equation}
\label{207}
    g_{E,10} = \left(\begin{array}{cc}
                  e^{2\Sigma} g_{E,4} & 0 \\
                      0 &  h
               \end{array}\right)\,,
\end{equation}
where $h$ is the six-dimensional internal metric and 
\begin{equation}
\label{208}
   e^{-2\Sigma} = \left(\det h\right)^{1/2}\,.
\end{equation}
This leads to the following four-dimensional lagrangian for metric and
 scalar fields:
\begin{equation}
\label{209}
   {\cal L}_{E,4} = \sqrt{|g|}\left[ R + {1\over 2}(\partial\phi)^2
    + 2 (\partial\Sigma)^2 -{1\over 4}\,{{\rm tr}}\,\partial h^{-1}\partial h
       \right]\,.
\end{equation}
We parametrize the internal metric as follows ($m,n=4,\cdots,9)$:
\begin{equation}
\label{210}
   -h_{mn} = e^{-\phi/2}\,{{\rm diag}}[
   e^{\chi/2},e^{\sigma/2},e^{\tau/2},e^{-\chi/2},e^{-\sigma/2},e^{-\tau/2}]\,.
\end{equation}
By construction, the solution for the four scalars in (\ref{210}) can
 be read off from
 the ten-dimensional solution (\ref{203}).

The reduction of the ten-dimensional gauge fields is straightforward.
 Here we only discuss the coupling of the resulting $d=4$ gauge fields
 to the scalars.
 The four harmonic functions $H_i$ contain four charges $Q_i$, which
 correspond to the solution for four $d=4$ $(p+2)$-form tensor 
 field-strengths. Therefore the reduction of the
 ten-dimensional $(p+5)$-form field-strength comprises three direct, and three
 double dimensional reductions. The solution (\ref{203},\ref{new}) tells us for
 which coordinates there is a double dimensional reduction:
 4,5,6 for $H_1$, 4,8,9 for $H_2$, 5,7,9 for $H_3$ and 6,7,8 for $H_4$.

The resulting $d=4$ action takes on the form:
\begin{eqnarray}
 &&  {\cal L}_{E,4} = \sqrt{|g|}\bigg[ R + 2(\partial\phi)^2
    + {1\over 8}\left( (\partial\chi)^2 + (\partial\sigma)^2
                        + (\partial\tau)^2 \right)  + \nonumber\\
  &&\quad {(-1)^{p+1}\over 2(p+2)!}
     e^{-2p\phi}\bigg( e^{-(\chi+\sigma+\tau)/2}\,F^2_{1,(p+2)} +
       e^{ -(\chi-\sigma-\tau)/2}\,F^2_{2,(p+2)} + \\
   \label{211}
  &&\quad   e^{ -(-\chi+\sigma-\tau)/2}\,F^2_{3,(p+2)} +
          e^{ -(-\chi-\sigma+\tau)/2}\,F^2_{4,(p+2)}\bigg) \bigg]\,.\nonumber
\end{eqnarray}
In the above, $F_{i,(p+2)}$ denotes the i-th field strength of rank $p+2$.
Note that in the case $p=2$ the four-index field-strengths can be
dualized to cosmological constants.

The solution of the equations of motion involving four scalar functions
 is:
\begin{eqnarray}
  ds^2_{E,4} &=& (H_1H_2H_3H_4)^{(p-1)/2}dx^2_{(p+1)} - \nonumber\\
\label{212}
  &&\qquad\qquad
                    (H_1H_2H_3H_4)^{(p+1)/2} dx^2_{(3-p)}\,,\\
\label{213}
  e^{\phi} &=& (H_1H_2H_3H_4)^{-p/4}\,, \\
\label{214}
  e^{\chi} &=& {H_3H_4\over H_1H_2}\,,\quad
  e^{\sigma} =   {H_2H_4\over H_1H_3}\,,\quad
  e^{\tau} =   {H_2H_3\over H_1H_4}\,.
\end{eqnarray} 
For simplicitly, we refrain from giving the expressions for the 
$(p+2)$-form field-strengths.

The multiscalar action (12) and the solution (\ref{212}-\ref{214})
 reproduce the solutions given in \cite{kh/lu/po,du/ra} (where also
 the solution for the gauge fields are given). 
Special cases are obtained by 
 setting one or more of the harmonic functions equal to unity. 
 In this way one finds solutions with one, two or three independent
 harmonic functions. If we identify the remaining harmonic functions,
 the resulting solutions can
 be expressed in terms of a single scalar $\varphi$, and the action takes 
 on the form
\begin{equation}
\label{216}
   {\cal L}_{E,4} = \sqrt{|g|}\left[ R + {1\over 2}(\partial\varphi)^2
      +{(-1)^{p+1}\over 2(p+2)!} e^{a\varphi} F^2_{(p+2)}\right]\,.
\end{equation}
Setting $N$ harmonic functions equal to each other and the
 remaining ones equal to one, result in four cases,
 $N=1,\ldots,4$, for which the constant $a$ in the action equals
\begin{equation}
    a = {1\over N}\sqrt{N^2(p^2-1) + 4N}
\end{equation}
and for which the solution reads:
\begin{eqnarray}
\label{217}
   && ds^2_{E,4} = H^{(p-1)N/2} dx^2_{(p+1)}
                        -H^{(p+1)N/2}dx^2_{(3-p)}\,,\nonumber\\
   && F_{0,1,\ldots,p,i} = \sqrt{N}\partial_i H^{-1}\,,
     \qquad
     e^{2\varphi} = H^{\sqrt{N^2(p^2-1) + 4N}}\,.
\end{eqnarray}
\noindent These results agree with the single harmonic
solutions obtained in \cite{lu/po/se/st}. The domain wall solutions
($p=2$) were also obtained in \cite{cv1} (for a recent review, see
\cite{cv2}).

\vspace{.5cm}
\noindent{\bf 3. Discussion}
\vspace{.5truecm}

In this letter we have reduced a ten-dimensional solution by
 standard Kaluza-Klein techniques to four dimensions. The 
 resulting solution
 can be interpreted as a configuration of four $p$-branes ($p=0,1,2$)
 in  four dimensions. The four-dimensional metric components
 are the overall world-volume and the overall transverse 
 components of the ten-dimensional metric.

Note however, that the four dimensional theory which we obtain 
 for $p=2$ contains four three-index gauge fields,
 or four independent cosmological constants
 after dualization of $F_{i,(4)}$. Clearly the $d=4$ lagrangian is
 then not a standard $d=4$ supergravity lagrangian\footnote{Since we start
 from the $d=10$ type II supergravity theories, one would expect
 $N=8$ supergravity in $d=4$ or truncations thereof.}. 
The three-index gauge fields arise from the Kaluza-Klein reduction 
of the $d=10$ six-index gauge field (which is the dual of the
IIB RR two-index gauge field).
 This is the source of the cosmological constants in the domain wall
 case.

Alternatively, one might have performed a duality
 transformation in $d=10$, turning the six-form gauge field into
 a two-index field. Then, in the reduction to $d=4$, no 3-form
 gauge fields appear, and, at first sight, no cosmological constants.
 However, now the reduction gives rise to additional scalars, which 
exhibit a shift symmetry that can be  used in a
 Scherk-Schwarz reduction \cite{sc/sc} to generate independent
cosmological constants.  This version of the 
 Scherk-Schwarz  technique was recently applied to the $d=10$ IIB theory
 \cite{be/ro/gr/pa/to}, and is extensively discussed, in the
 context of domain walls in diverse dimensions, in
 \cite{co/lu/po/st/to}. Cosmological constants can also be obtained
 directly in  supergravity theories by suitably chosen gaugings of its global
 symmetries \cite{sa/se}. In fact, it turns out that
the results of the Scherk-Schwarz
 reduction can be recaptured by such gaugings in lower
 dimensions \cite{be/ro/ey}. 
 
It is natural to generalize the results of the present work to
six dimensions. 
 To obtain $d=6$ solutions with a similar structure we can only use
 intersections of two $D$-branes in $d=10$ \cite{be/be/ja}. The intersection of
 two $(p+2)$-branes in $d=10$ gives rise to $p$-brane solutions
 in $d=6$ with $p=0,\ldots,4$. In this case the lagrangian is
\begin{eqnarray}
   {\cal L}_{E,6} &=& 
   \sqrt{|g|}\bigg[ R + (\partial\phi)^2
    + {1\over 4}(\partial\chi)^2 +  \nonumber\\
   \label{321}
 &&\qquad
  e^{-(p-1)\phi}\bigg( e^{-\chi}\,F^2_{1,(p+2)} +
       e^{ \chi}\,F^2_{2,(p+2)} \bigg)\bigg]\,. 
\end{eqnarray}
The solution for the metric and scalars is
\begin{eqnarray}
 &&  ds^2_{E,6} = (H_1H_2)^{(p-3)/4}dx^2_{(p+1)} 
         - (H_1H_2)^{(p+1)/4}dx^2_{5-p}\,, \nonumber\\
 &&   e^{\phi} = (H_1H_2)^{(1-p)/4}\,,\quad
      e^\chi = {H_1\over H_2} \,.
\end{eqnarray}
 In this case the domain wall solution, $p=4$, is associated with
 a seven-index gauge field in the $d=10$ IIA theory, the dual of
 the RR vector. 

So far we have used $T$-duality in a rather specific way to obtain 
 intersections of identical $D$-branes in $d=10$. By additional 
 duality transformations we can also reach from, e.g, \fiveint\ ,
 the intersections \foursixint,\ \foureightint,\ \twosixint,\ and
 \threefivesevenint. Although the solution is different in $d=10$, 
 the
 reduction and truncation leads to the same solution 
 in $d=4$ up to redefinitions of the scalar fields.

In some cases the reduced action has an enhanced $SL(2,\R)$ symmetry. 
This is the
 case for $p=1$ in $d=4$ ($p=3$ in $d=6$), 
 if three (one) harmonic functions are set equal to 
 unity. In both $d=4$ and $d=6$ a duality transformation which
 turns the gauge field into a scalar is required to make the
 $SL(2,\R)$ symmetry explicit in the action. 
 These solutions can be obtained by
 double dimensional reduction from the seven-brane in the IIB theory.
 This couples to an eight-index gauge field, which is dual to the
 IIB RR scalar. Thus the lower dimensional $SL(2,\R)$ symmetry is an
 immediate consequence of the IIB $SL(2,\R)$ dilaton/RR-scalar symmetry.
 
Finally,
the solutions we have obtained with four harmonic functions have
 two unbroken supersymmetry generators. It would be interesting to 
 understand the supersymmetry of these solutions from the 
 four-dimensional point of view. In particular, one would like to know
 in which supergravity theory ($N=4, N=8$) the different
solutions can be embedded. 
 This requires an extension of
 the analysis of \cite{kh/or} to the case of four-dimensional strings and
 domain walls.
 
\vspace{.5truecm}
\noindent {\bf Acknowledgements}
\vspace{.5truecm}

This work is part of the research program of the ``Stichting
 voor Fundamenteel Onderzoek der Materie'' (FOM). 
 It is also supported  by the European Commission TMR programme 
 ERBFMRX-CT96-0045,
 in which E.B. and M.~de R. are associated to the University of Utrecht.
 S.~Panda thanks the
 FOM for financial support, and the Institute of Theoretical Physics
 in Groningen for its excellent hospitality.


\vspace{.5truecm}


\begin{thebibliography}{99}
\bibitem{du/kh/lu}
M.~J.~Duff, R.~R.~Khuri and J.~X.~Lu, 
Phys.~Rep. 259 (1995) 213.
\bibitem{ra}
J.~Rahmfeld, 
%{\it Extremal black holes as bound states},
%{\tt hep-th/9512089}.
Phys.~Lett.~B372 (1996) 198.
\bibitem{cv/ts}
M.~Cvetic and A.~A.~Tseytlin,
{\it Solitonic strings and BPS saturated dyonic black holes},
{\tt hep-th/9512031}.
\bibitem{kh/lu/po} N.~Khviengia, Z.~Khviengia, H.~L\"u and
C.N.~Pope, {\it Intersecting $M$-branes and Bound States}, 
{\tt hep-th/9605077}.
\bibitem{du/ra}
M.~J.~Duff and J.~Rahmfeld,
{\it Bound states of black holes and other $p$-branes},
{\tt hep-th/9605085}.
\bibitem{pa/to}
G.~Papadopoulos and P.~K.~Townsend, 
%{\it Intersecting $M$-branes},\hfill
%{\tt hep-th/9603087}.
Phys.~Lett.~B380 (1996) 273.
\bibitem{ts1}
A.~A.~Tseytlin,
{\it Harmonic superpositions of M-branes}, 
{\tt hep-th/9604035}; 
I.~R.~Klebanov and A.~A.~Tseytlin,
{\it Intersecting $M$--branes as four-dimensional black holes},
{\tt hep-th/9604166}.
\bibitem{be/be/ja}
K.~Behrndt, E.~Bergshoeff and B.~Janssen,
{\it Intersecting $D$--Branes in Ten and Six Dimension},
{\tt hep-th/9604168}.
\bibitem{ga/ka/tr} 
J.~P.~Gauntlett, D.~A.~Kastor and J.~Traschen,
{\it Overlapping Branes in $M$--Theory},
{\tt hep-th/9604179}.
\bibitem{ba/la}
V.~Balasubramanian and F.~Larsen, 
{\it On D-Branes and Black Holes in Four Dimensions}, 
{\tt hep-th/9604189}.
\bibitem{be/be}
K.~Behrndt and E.~Bergshoeff,
{\it A Note on Intersecting $D$-branes and Black Hole Entropy},
{\tt hep-th/9605216}.
\bibitem{ts2}
A.~A.~Tseytlin, 
{\it Composite black holes in string theory},
talk at the Second International Sakharov Conference in Physics,
 Moscow 1996, 
{\tt gr-qc/9608044}.
\bibitem{lu/po}
H.~L\"u and C.~N.~Pope,
{\it Multi-scalar $p$-brane solutions},
{\tt hep-th/9512153}.
\bibitem{po}
J.~Polchinski, Phys.~Rev.~Lett.~{\bf 75} (1995) 184.
\bibitem{po/ch/jo}
J.~Polchinski, S.~Chaudhuri and C.~V.~Johnson,
{\it Notes on D-branes},
{\tt hep-th/9602052}.
\bibitem{be/hu/or}
E.~Bergshoeff, C.~M.~Hull and T.~Ort\'{\i}n,
Nucl.~Phys.~B451 (1995) 547.
\bibitem{ber/roo} E.~Bergshoeff and M.~de Roo, Phys.~Lett.~B380 (1996) 265.
\bibitem{be/bo/or}
E.~Bergshoeff, H.~J.~Boonstra and T.~Ort\'{\i}n,
Phys.~Rev.~D53 (1996) 7206.
\bibitem{lu/po/se/st}
H.~L\"u, C.~N.~Pope, E.~Sezgin and K.~S.~Stelle,
Nucl.~Phys.~B456 (1995) 669.
\bibitem{cv1} 
M.~Cvetic, Phys.~Lett.~B341 (1994) 160.
\bibitem{cv2} 
M.~Cvetic and H.H.~Soleng, {\it Supergravity domain walls},\hfill\break
{\tt hep-th/9604090}.
\bibitem{sc/sc}
J.~Scherk and J.~H.~Schwarz,
Phys.~Lett.~B82 (1979) 60.
\bibitem{be/ro/gr/pa/to}
E.~Bergshoeff, M.~de Roo, M.~B.~Green, G.~Papadopoulos and P.~K.~Townsend,
Nucl.~Phys.~B470 (1996) 113.
\bibitem{co/lu/po/st/to}
P.~M.~Cowdall, H.~L\"u, C.~N.~Pope, K.~S.~Stelle and P.~K.~Townsend,
{\it Domain walls in Massive Supergravities},
{\tt hep-th/9608173}.
\bibitem{sa/se}
A.~Salam and E.~Sezgin,
{\it Supergravities in diverse dimensions},
North-Holland/World Scientific (1989).
\bibitem{be/ro/ey}
E.~Bergshoeff, M.~de Roo and E.~Eyras,
in preparation.
\bibitem{kh/or}
R.~R.~Khuri and T.~Ort\'{\i}n,
Phys.~Lett. B373 (1996) 56;
R.~R.~Khuri and T.~Ort\'{\i}n,
Nucl.~Phys.~B467 (1996) 355.

\end{thebibliography}
\end{document}